\AtBeginDocument{\RenewCommandCopy\qty\SI}  
\documentclass[a4paper,fleqn]{cas-dc}
\usepackage{amsmath, amssymb, wasysym}
\usepackage{physics}
\usepackage{siunitx}
\usepackage{lineno}
\usepackage{dcolumn}
\begin{document}
\def\floatpagepagefraction{1}
\def\textpagefraction{.001}
\shorttitle{Force-balanced Halbach magnets}
\shortauthors{I.~Rehberg et~al.}

\title [mode = title]{Openable Force-Balanced Halbach Magnets: From Fibonacci Sphere Simulations to Icosahedral Realizations}

\author[1]{I. Rehberg}[orcid=0000-0003-1547-2951]
\credit{Theory, Calculations, Software, Vizualization, Writing}
\affiliation[1]{organization={Institute of Physics, University of Bayreuth}, city={Bayreuth},postcode={95440}, country={Germany}}

\author[2]{H. Soltner}[orcid=0000-0001-8595-3393]
\credit{Theory, Calculations, Writing}
\affiliation[2]{organization={Institute of Technology and Engineering (ITE), Forschungszentrum J\"ulich GmbH}, city={J\"ulich},postcode={52425}, country={Germany}}

\author[3]{P. Bl\"umler}[orcid=0000-0002-0412-7549]
\cormark[1]
\credit{Coordination, Writing, Original draft preparation, Construction of Prototypes, Measurements}
\affiliation[3]{organization={Institute of Physics, University of Mainz}, city={Mainz},postcode={55128}, country={Germany}}
\cortext[cor1]{Corresponding author}
\begin{abstract}
A long-standing goal in magnet design is to completely surround a volume of highly homogeneous magnetic field with permanent magnets while maintaining practical access to that volume. In this work, we present a theoretical and experimental investigation of mechanically accessible spherical magnets in Halbach configuration that can be opened with minimal or vanishing force. Focusing on dipolar Halbach spheres composed of discrete magnetic subunits, we derive conditions for force-free opening along specific cutting planes. These conditions define a continuous set of geometries for which tensile magnetic forces cancel, leaving only shear components, enabling mechanically effortless opening. The theoretical predictions are validated experimentally using icosahedral approximations of the Halbach sphere, for which both opening forces and magnetic field properties are measured. The results demonstrate that excellent field homogeneity can be preserved while reducing opening forces by orders of magnitude. Although discussed in detail for the dipolar case, the theoretical framework is general and applicable to higher-order multipole Halbach systems. Finally, the concepts are extended to spherocylindrical Halbach configurations, highlighting their potential for large-volume, highly homogeneous, and mechanically accessible permanent-magnet systems for magnetic resonance and related applications.
\end{abstract}

\begin{graphicalabstract}
\includegraphics[width=.9\textwidth]{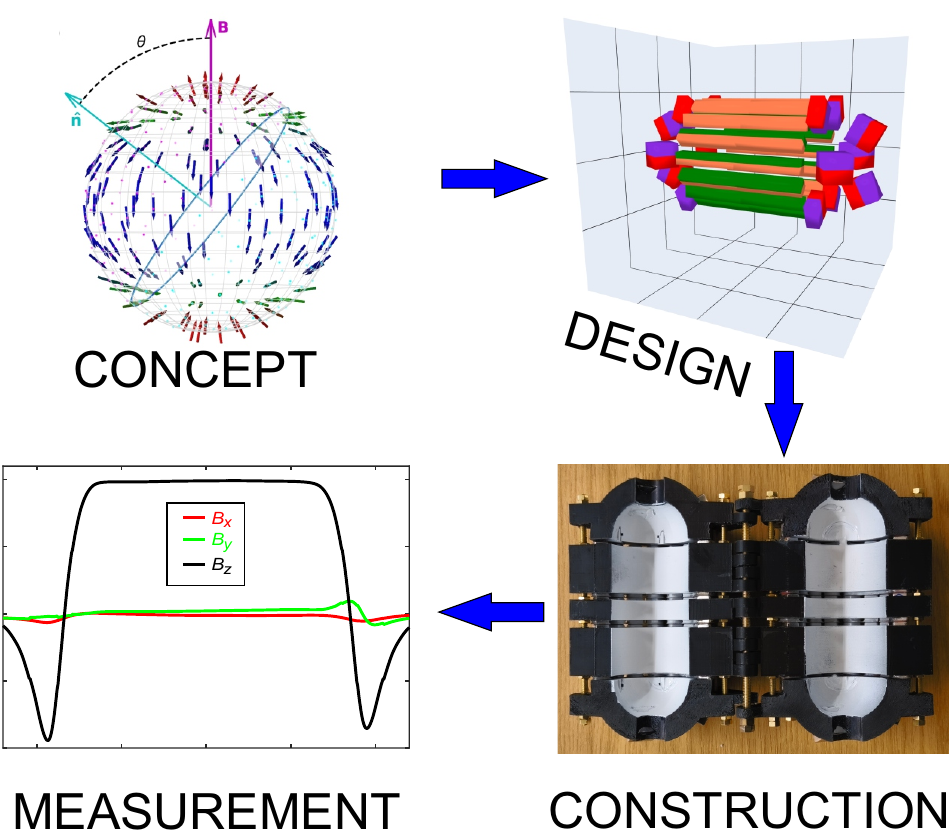}
\end{graphicalabstract}

\begin{highlights}
\item Numerically computed cutting planes enable force-free opening of discrete spherical Halbach magnets.
\item The theoretical framework is general and directly applicable to higher-order multipole Halbach configurations.
\item Icosahedral arrangement of magnets demonstrate reduced opening forces while preserving magnetic field homogeneity.
\item Spherocylindrical geometries enable scalable systems with larger, expandable accessible volumes of highly homogeneous fields.
\item Very efficient design in terms of magnetic material usage relative to the generated homogeneous field volume.
\item Overcomes the limited accessibility of spherical magnets, enabling general applications.
\item Enables compact, energy-efficient permanent-magnet systems for magnetic resonance and related applications.
\end{highlights}

\begin{keywords}
Permanent magnets \sep Magnetic resonance \sep Halbach magnets \sep Homogeneous magnetic fields \sep Fibonacci sphere \sep Icosahedral symmetry \sep Magic-angle spinning 
\end{keywords}
\maketitle  

\section{Introduction}
Magnetic resonance techniques such as nuclear magnetic resonance (NMR) and electron spin resonance (ESR) rely on the application of strong, spatially homogeneous static magnetic fields. In NMR in particular, magnetic field homogeneity directly limits spectral resolution, while field strength determines both sensitivity and achievable resolution. Although superconducting magnets dominate high-field laboratory systems, permanent magnets have become increasingly attractive for specialized applications in recent years \cite{Johns2015}.

Permanent magnets offer two decisive advantages. First, they deliver exceptionally high magnetic field strength per unit volume: modern rare-earth materials (e.g., NdFeB) can produce flux densities comparable to those generated by coils requiring thousands of ampere-turns and, in the same confined volume, incurring substantial resistive losses ---  \mbox{often} on the order of hundreds of watts. In contrast, permanent magnets operate entirely passively, requiring neither electrical power nor active cooling to maintain the field. Together, these characteristics enable compact, energy-efficient and portable magnetic resonance systems suitable for mobile and field-deployable applications \cite{Johns2015}.

Achieving strong and homogeneous magnetic fields with permanent magnets requires carefully designed arrangements that optimally superimpose the fields of individual magnet elements. The Halbach array \cite{Halbach1980} provides an elegant and well-established solution to this problem. There is a great deal of literature on the design and construction of Halbach cylinders, as summarized in a recent review \cite{Soltner2023}.

In an ideal realization, a Halbach magnet would be fabricated from a magnetic material with continuously varying magnetization directions \textendash a spatial waveform whose frequency determines the polarity and distribution of the resulting field. Since such materials are not readily manufacturable, practical implementations rely on discretization into multiple magnetic subunits, each with a uniform magnetization direction, arranged to optimize field strength and/or homogeneity \cite{RehbergBluemler2025}. This modular construction not only approximates the ideal field distribution, but also allows, in principle, relative motion between the discrete subunits.

This relative motion has been exploited in several ways. One approach uses the rotation of two or more concentric Halbach assemblies to add or subtract magnetic fields, thus tuning the overall field strength \cite{ Soltner2023, Leupold1993, Bauer2009, Tretiak2019}. Another possibility is to mechanically open the magnet structure, for example, by means of a hinge, which ideally requires no force \cite{CUFF2011}. In that publication, and subsequently in \cite{Soltner2023}, force-free angles were analyzed for cylindrical Halbach arrangements composed of magnetic dipoles.

Shortly thereafter, similar concepts were proposed for spherical Halbach configurations \cite{patent_spheres,BluCasa2015}. However, only recently has it become possible to realize spherical Halbach magnets approximated by regular (Platonic) and semi-regular (Archimedean) solids \cite{RehbergBluemler2026}. That work demonstrated that the highest magnetic field homogeneity is associated with icosahedral symmetry.

Despite their excellent field properties, the practical use of spherical Halbach magnets has long been considered severely constrained, because of the limited access to their interior or the large magnetic forces encountered when opening the structure. Depending on geometry and magnet strength, opening forces can reach several kilonewtons, posing substantial challenges in handling, safety, and mechanical design. Previous attempts to mitigate this problem have included introducing apertures or slots into the magnet structure or superimposing magnetized regions with spatially varying remanence \cite{Leupold1994, Leupold2000}. Although these approaches can provide limited access, they generally compromise field homogeneity, increase stray fields and weight, or require complex and impractical fabrication processes.

In the present work, the mechanical opening of discretized spherical Halbach magnets is investigated both theoretically and from a practical perspective. Although the discussion focuses on the dipolar Halbach sphere, the underlying theory is general and applies equally to higher-order multipole magnet systems. Guided by the theoretical predictions, icosahedral magnet arrangements were constructed, and both the forces required to open them and the resulting magnetic field properties are experimentally characterized.

\begin{figure*}[!hbtp]
\centering
\includegraphics[width=\textwidth]{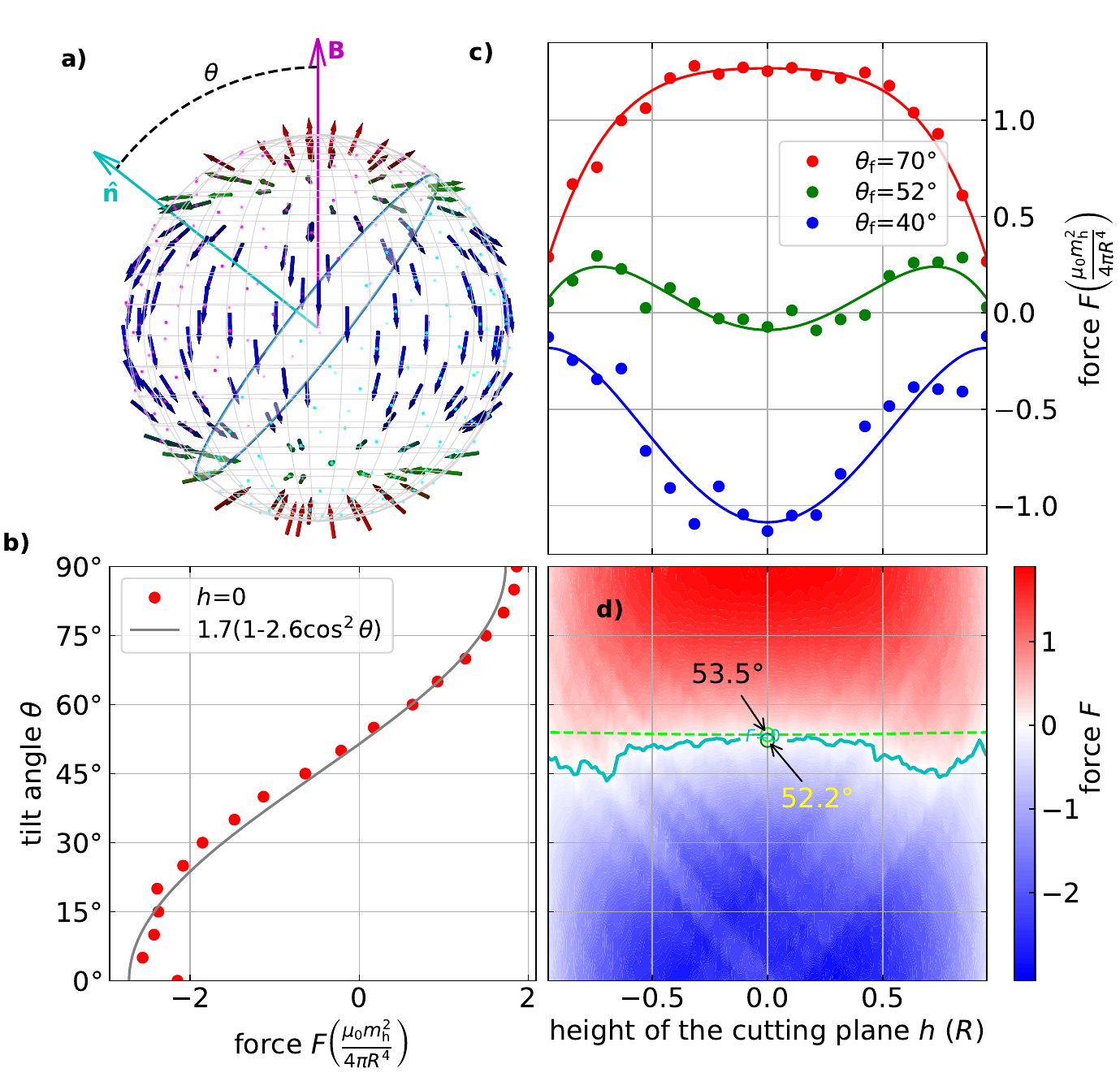}
\caption{The force to open a spherical Fibonacci arrangement of 300 dipoles. (a)~Only every second dipole is indicated by an arrow pointing along $\mathbf{m}$. Their arrangement forms a homogeneous field indicated by  $\mathbf{B}$ (magenta), where red arrows are almost parallel to $\mathbf{B}$, blue antiparallel, green perpendicular. The vector $\mathbf{\hat{n}}$ (cyan) is normal to the cutting plane (cyan circle).  (b)~$F(\theta, h=0)$ is numerically determined (red dots). The parameters of the fitted solid gray line are given in the legend. (c) $F(\theta_\text{f},h)$ for three different angles $\theta_\text{f}$: attracting (blue), almost forceless (green), repelling (red). The lines are smoothing splines to guide the eye. (d) $F(\theta, h)$: the green solid line denotes $F\!=\!0$. The zero-force line for 20,000 dipoles is dashed green.}  
\label{fig:1}
\end{figure*}

Finally, the concepts developed here are extended to a combination of Halbach cylinders and hemispheres that form a spherocylindrical magnet (capsule-shaped) \cite{Soltner2023, Chen2007}. In principle, such magnets can be fabricated with arbitrary lengths and are therefore capable of enclosing extended objects (e.\ g., humans) while providing highly homogeneous magnetic fields over a large fraction of the internal volume. Access to the interior of such a magnetic “sarcophagus” can be easily achieved by opening the structure at an appropriate angle, ideally requiring only minimal force \textendash or even exploiting a controlled repelling force to compensate for the weight of the moving section.

This represents something of a magnet designer’s ideal: a volume of highly homogeneous magnetic field, fully enclosed by magnets yet entirely accessible, and maximally efficient in terms of magnet mass relative to the volume of the homogeneous field.

\section{Theory}
\subsection{Separation force of discretized Halbach spheres}
As real Halbach spheres will always feature a degree of discretization with smaller or larger permanent magnets, the basic concept of a force-free cut can be illustrated by considering an arrangement of a sufficiently large number of point dipoles located on a sphere with radius $R$.

As an example, an arrangement of $N\!=\!300$ dipoles distributed on the surface of a sphere is shown in Fig.~\ref{fig:1}a. The orientation of the field in the center of the sphere is indicated by $\mathbf{B}$ and is oriented along the z-axis. The normal of the cutting plane,
\begin{equation*}
\mathbf {\hat{n}}=
\begin{pmatrix}
\sin\theta\,\cos\phi\\
\sin\theta\,\sin\phi\\
\cos\theta
\end{pmatrix},
\end{equation*}
forms the polar angle $\theta$ with $\mathbf{B}$, while the azimuthal angle $\phi$ is $180^\circ$ in this drawing. The cutting plane is allowed to be shifted along $\mathbf{\hat{n}}$, where $h$ is the distance of the cutting plane from the center of the sphere. 

To calculate forces between the magnets on different sides of the cutting plane we recall that the force $\mathbf{F}$ between two magnetic dipoles with magnetic moments $\mathbf{m}_1$ and $\mathbf{m}_2$, separated by the vector $\mathbf{r}$, is given by
\begin{equation}
\begin{split}
\mathbf{F}(\mathbf{r})\, &= \,
\nabla \, \left( \mathbf{m}_1 \, \frac{\mu_0}{4\pi} \,\frac{3(\mathbf{m}_2 \cdot \mathbf{r})\mathbf{r} - \mathbf{m}_2 r^2}{r^5} \right)\\[5pt]
  &= \, \frac{\mu_0}{4\pi} \, \nabla \, \left( \frac{3(\mathbf{m}_1 \cdot \mathbf{r}) (\mathbf{m}_2 \cdot \mathbf{r})}{r^5}  -  \frac{(\mathbf{m}_1 \cdot \mathbf{m}_2)}{r^3} \right),\\
\end{split}
\label{eq:1_Fdip}
\end{equation}
where $r = |\mathbf{r}|$ and $\mu_0 \approx 4\pi \cdot 10^{-7} \,\text{Vs}/\text{Am}$ denotes the permeability of free space \cite{Furlani2001}. Applying the gradient operator then gives (for example, with $\hat{\mathbf{m}}_1$ as the unit vector of $\mathbf{m}_1$ and $m_1 \!=\! |\mathbf{m}_1|$ as its magnitude) \cite{Yung1998}
\begin{equation}
\begin{split}
\mathbf{F}(\mathbf{r})\, = &\, \frac{3 \mu_0 m_1 m_2 }{4 \pi r^4} \left\{ (\hat{\mathbf{m}}_1 \cdot \hat{\mathbf{r}}) \hat{\mathbf{m}}_2 + (\hat{\mathbf{m}}_2 \cdot \hat{\mathbf{r}}) \hat{\mathbf{m}}_1 -  \right.\\[5pt]
& \left. \left[5(\hat{\mathbf{m}}_1 \cdot \hat{\mathbf{r}}) (\hat{\mathbf{m}}_2 \cdot \hat{\mathbf{r}}) - (\hat{\mathbf{m}}_1 \cdot \hat{\mathbf{m}_2})   \right]\hat{\mathbf{r}} \right\} .\\
\end{split}
\label{eq:2}
\end{equation}

The force between the two clusters separated by the cutting plane can then be obtained as the sum of such dipole-dipole interactions, and the result is illustrated in Fig.~\ref{fig:1}. Here we consider $N\!=\!300$ dipoles located on a Fibonacci lattice \cite{Swinbank2006}, which provides an approximately uniform equal-area distribution of points on the sphere. Interactive explorations of its magnetic properties are provided in \cite{RehbergBluemler2026a,Rehberg2026b}. The locations of the dipoles are shown as dots, cyan below the cutting plane, and magenta above. They are oriented according to the Halbach condition, namely that the polar orientation angle is twice the polar angle of its location. Every second dipole orientation is indicated as an arrow.

In addition, an example of a cutting plane is shown. The distance $h$ of this plane from the center of the sphere is measured in units of its radius $R$, and $h=0R$ in this drawing. 

An overview of the force component $F\!=\!\mathbf{F}(\theta, h)\cdot \hat{\mathbf{n}}(h)$ between the two spherical caps separated by the cutting plane is given in Fig.~\ref{fig:1}d. Attractive forces are shown in blue and repelling forces are shown in red. For a qualitative understanding of the sign change of the force, it helps to consider the line along $h\!=\!0R$ and to interpret each half sphere as a single dipole. For $\theta \!=\! 0^\circ$ we are dealing with 2 parallel dipoles located one behind the other - this yields an attractive force. For $\theta \!=\! 90^\circ$ the two parallel dipoles are located side by side - they repel each other. The points of interest, where the force is zero, must lie somewhere between these extremes. They are indicated by the solid green line. The kinks in that line occur whenever the cutting plane crosses a (point) dipole. The dashed line denotes zero force for 20,000 magnets, which can be considered as an approximation of a continuous distribution of magnetization, an argument which is elucidated in further detail below.

The line along $h\!=\!0R$ is shown in quantitative detail in Fig.~\ref{fig:1}b. The force $F$ is given in units of
\begin{equation}
\frac{\mu_0 m_\text{h}^2}{4\pi R^4}, \quad \text{with}\quad  m_\text{h}=\frac{N}{2}m.
\label{eq:3_force_unit}
\end{equation}
 Thus, $m_\text{h}$ is the sum of all the magnitudes of dipole moments located on the surface of a hemisphere. By this scaling, one obtains numbers for the force that are fairly independent of the number of dipoles considered. The numerically calculated forces are indicated by red dots. For comparison, a curve 
 \begin{equation*}
 F\propto 1-\lambda \,cos^2 \theta
 \end{equation*}
 is also shown in the plot, where $\lambda$ is a fit parameter. This is a purely empirical ansatz, inspired by the force between two parallel dipoles, where $\lambda$ would be 3. The fact that we are getting 2.6 instead indicates that the simplified model of two point dipoles interacting at a fixed distance -- which does explain the qualitative features -- is not sufficient for a quantitative description.

Figure~\ref{fig:1}c shows three line cuts through the contour shown in Fig.~\ref{fig:1}d. At a cutting angle of $\theta\!=\!40^\circ$, we have an attractive force, which becomes weaker with increasing distance of the cutting plane from the center. For $\theta\!=\!52^\circ$ the force is still below zero for $h\!=\!0R$, but slightly above for larger values of $\lvert h\rvert$. At $\theta\!=\!70^\circ$, the force is clearly repulsive, becoming weaker with increasing $\lvert h\rvert$. The solid smoothing splines in this plot are just intended as guides for the eye. Note that the data points are not perfectly symmetric with respect to the $h\!=\!0R$ line because the Fibonacci distribution of the dipoles does not have exact point symmetry with respect to the center of the dipole cluster. 

As shown in Fig.~\ref{fig:1}d, the force is much more sensitive to changes in the tilt angle $\theta$ than to variations in the height of the cutting-plane $h$, but the tendency that the force is reduced with increasing $\lvert h\rvert$ -- a reduced opening aperture -- is visible.

As a side remark on the Fibonacci distribution of permanent magnets, it is noticeable that it is also a practical approach to achieve homogeneous magnetic fields inside a sphere. Although it does not offer a clear symmetry of the magnetic field around the center and is thus inferior to the fourth-order saddle points provided by the configurations with icosahedral symmetry, it can outperform those configurations once a finite deviation from the center field is accepted. To give an example: The volume with a 1\% deviation from the center is larger for a Fibonacci arrangement of 120 magnets than for a distribution of those magnets at the vertices of a truncated icosidodecahedron. However, the latter arrangement wins if only deviations on a ppm-level are acceptable \cite{RehbergBluemler2026a}.
\begin{figure*}[!htbp]
\centering
\includegraphics[width=\textwidth]{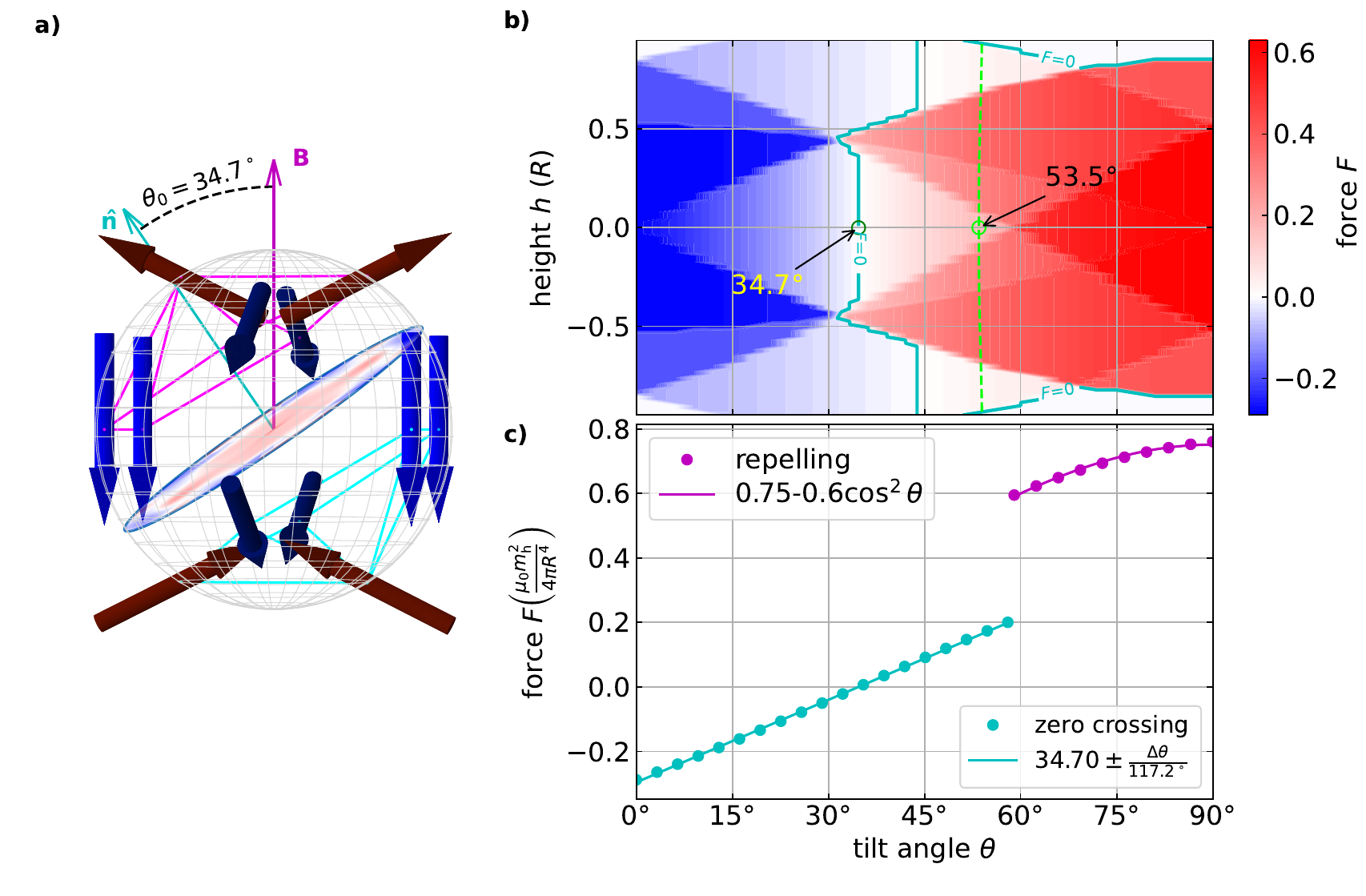}
\caption{The force to open an icosahedron. (a) Same as Fig.~\ref{fig:1}a, for 12 magnets in an icosahedral arrangement. (b) Same as Fig.~\ref{fig:1}d. (c) Same as Fig.~\ref{fig:1}b. $\Delta\theta\!=\!\theta-\theta_0$.}
\label{fig:2}  
\end{figure*}

A smaller number of magnets is clearly more practical than the 300 dipoles shown above.  12 magnets, located at the vertices of an icosahedron, and oriented in the spherical Halbach arrangement, form a very homogeneous magnetic field in their center, precisely speaking a saddle point of fourth order  \cite{RehbergBluemler2026}. Thus, they constitute a very effective and useful spherical arrangement that merits investigation of its opening force. The geometry is shown in Fig.~\ref{fig:2}a. Note that the position of the vertices is chosen to preserve symmetry with respect to the $xy$-, $xz$- and $yz$-plane. 

The opening force $F(\theta, h)$ for this configuration is shown in Fig.~\ref{fig:2}b. The fact that this is a function with jump discontinuities is much more prominent here compared to the Fibonacci arrangement with 300 magnets given above. The most practicable opening plane is certainly the one with $h\!=\!0R$, a great circle that cuts the sphere exactly into two hemispheres. It should be noted that the angle of a force-free opening is smaller here compared to the  configuration discussed above, namely $\theta_0\approx 34.7^\circ$. The force near this angle crosses the zero line with a finite slope of  about $1/117^\circ$ as shown in Fig.~\ref{fig:2}c. The unsteady jump near $\theta\!=\!60^\circ$ is caused by the fact that two dipoles exchange their side with respect to the cutting plane here. 

\begin{figure*}[!htbp]
\centering
\includegraphics[width=\textwidth]{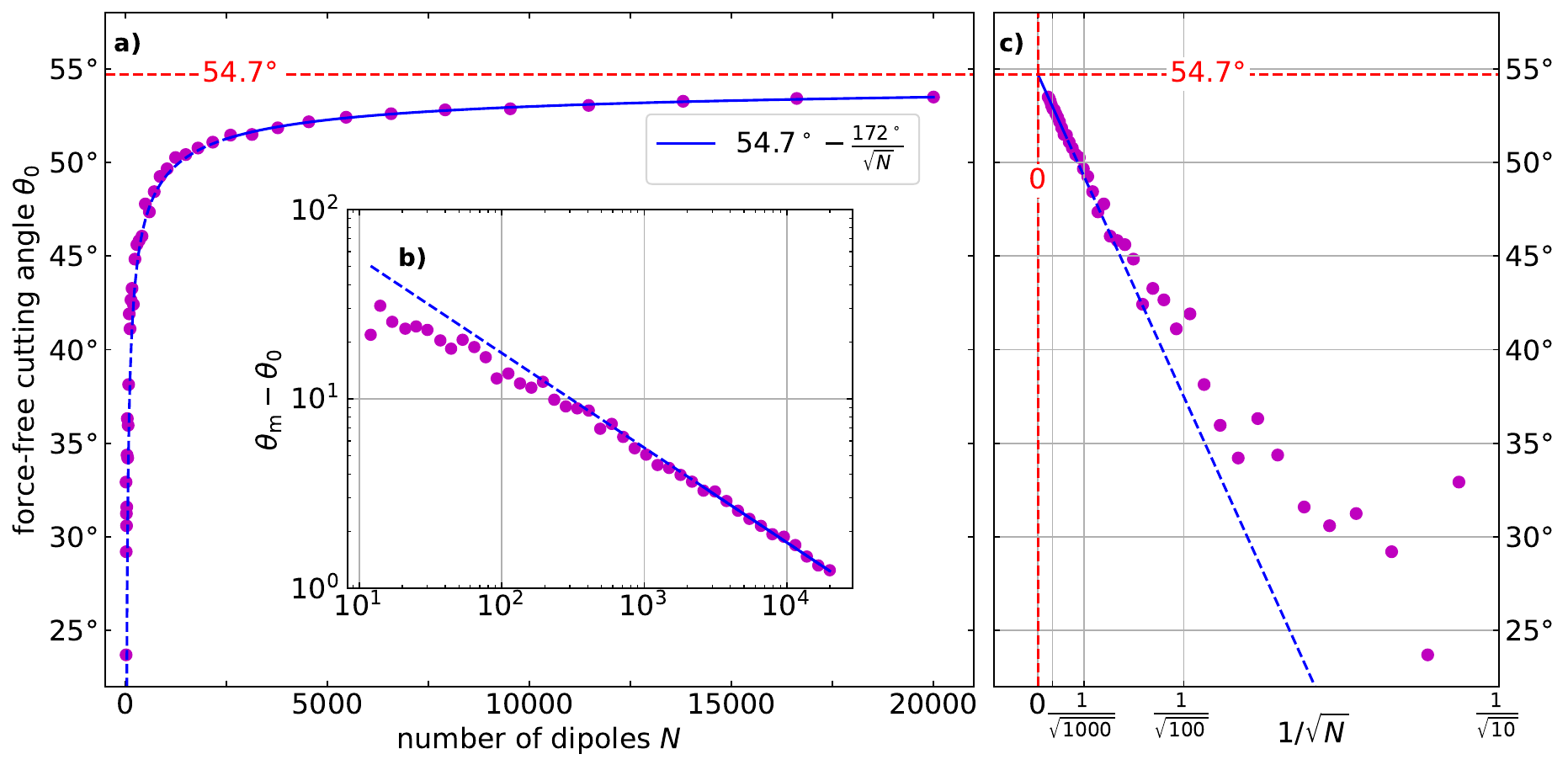}
\caption{The opening angle for Fibonacci spheres increases with the number of dipoles. (a) The dots (magenta) are numerically obtained. The fitting function (blue) with two parameters is given in the legend. The fitting range is indicated by a solid line, and the extrapolated part is dashed. The asymptotic value obtained from that fit is indicated as the red dashed line. (b) This inset shows $\theta_\text{m}-\theta_\text{0}(N)$ on a log-log plot. (c)~Same data shown as function of $1/\sqrt{N}$.   In this case, $\lim_{N\to\infty}\theta_0(N)=\theta_\text{m}$ is revealed as the crossing of the asymptotic straight line with the line $x\!=\!0$.}
\label{fig:3}
\end{figure*}

Having noticed that the force-free opening angle increases to $\theta_0\!=\!52.2^\circ$ for 300 dipoles, and to $\theta_0\!=\!53.5^\circ$ for 20000 dipoles,  one can speculate that it reaches the magic angle in the limit of a homogeneous Halbach magnetization of the sphere, although there is no mathematical proof  at this point, instead we can only present numerical evidence in Fig.~\ref{fig:3}.

Figure~\ref{fig:3} shows that the cutting angle $\theta_0$ increases with increasing number of dipoles. Although one might think that a distribution of 20000 vertices is already a very good approximation of a homogeneous Halbach distribution of magnetization, the plot clearly shows that we are still dealing with an error on a percentage level here. Specifically speaking, the deviation of the cutting angle from the asymptotic one according to the fit is $172^\circ/\sqrt{20000}\approx 1.2^\circ$.

Figure Fig.~\ref{fig:3}b (the inset of Fig.~\ref{fig:3}a) confirms that the asymptotic range is reached for $N>500$, and shows the power -0.5 by the asymptotic slope in this log-log plot. Once this power is known, the asymptotic cutting angle can also be determined graphically, by plotting $\theta_0$ as a function of $1/\sqrt{N}$, as done in Fig.~\ref{fig:3}b. The asymptotic value is read from the crossing of $\theta_0(1/\sqrt{N})$ with the zero line. 

We take the study shown in Fig.~\ref{fig:3} as convincing evidence that the asymptotic value of the cutting angle is the magic angle:
\begin{equation}
 \lim_{N\to\infty}\theta_0(N)=\theta_\text{m}=\arccos\left({1}/{\sqrt{3}}\right)
 \end{equation}
This result is consistent with interpreting both hemispheres as point dipoles of equal strength and parallel magnetization. According to Eq. ~\ref{eq:2}, their force would depend on the cutting angle $\theta$ like $F\propto 1-3\cos^2\theta$. This term crosses zero at the magic angle. However, it must be noted that this argument might be useful for memorizing the result, but it is not exact because it ignores the forces that arise from higher-order moments of the hemisphere.

A technical note on the numerical summation of forces is useful here. For the example shown in Fig.~\ref{fig:1}a, there are $150\times150$ pairs to add. If only one dipole is left on one side of the cutting plane, by choosing a sufficiently large $h$, the number of interactions reduces to 299 in this particular case. A direct summation of forces is certainly possible for a small number of dipoles, but it might cause numerical problems for larger numbers. The largest number of dipoles we considered here is 20000, as shown in Fig.~\ref{fig:3}. In that case $10^8$ pairs have to be added, where the largest distance is 2 and the smallest distance about 0.02, thus their distances differ by two orders of magnitude. However, due to the strong dependence on distance, namely $F\propto r^{-4}$, this means adding up $10^8$ numbers, where the force terms differ by 8 orders of magnitude. This problem can be reduced slightly by summing up potentials, the term within the bracket in eq.~\ref{eq:1_Fdip},  because the potential drops with only the third power of the distance. The gradient is then obtained numerically by the difference of two potentials calculated at a small distance.

\subsection{Force-Balanced Halbach spherocylinders}
\begin{figure*}[!htbp]
\centering
\includegraphics[width=0.8\textwidth]{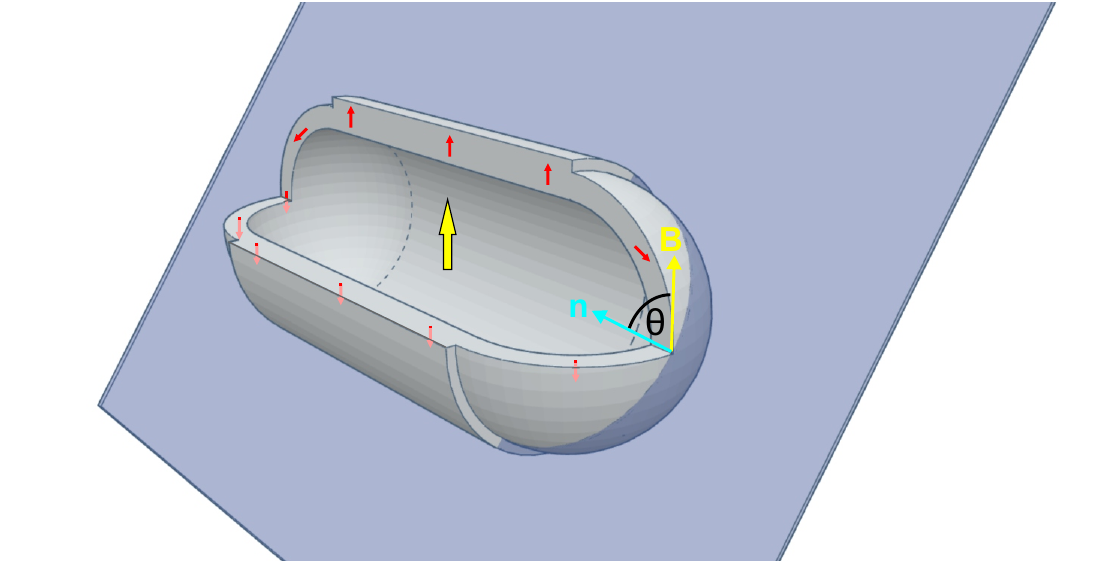}
\caption{Halbach hemispheres as end caps of a Halbach cylinder form a Halbach spherocylinder (capsule shape). A front quarter is cut away to reveal the interior. The red arrows indicate the magnetization directions, and the large yellow arrow denotes the resulting magnetic field. The blue transparent plane marks the surface for an opening at an angle of $\theta \!=\! \theta_{\text{m}}$ between the normal $\hat{\mathbf{n}}$ (cyan) and the magnetic field $\mathbf{B}$ (yellow).}
\label{fig:4}
\end{figure*}

Spherocylinders are Halbach cylinders closed by  spherical caps to minimize the distortions of the homogeneous field at their ends, as described in~\cite{Soltner2023, Chen2007}, and illustrated in Fig.~\ref{fig:4}.
To find the appropriate radius for the end caps, it helps to recall that an ideal Halbach spherical shell  with an inner radius $^\text{s}R_\text{i}$ and an outer radius $^\text{s}R_\text{o}$ generates a magnetic field of \cite{Leupold2000} 
\begin{equation*}
B_\text{c} = B_\text{R} \frac{4}{3} \ln\left( \frac{^\text{s}R_{\text{o}}}{^\text{s}R_{\text{i}}}\right)
\end{equation*}
that is 4/3 stronger than that of an infinitely long  Halbach cylinder with the same inner and outer radii ( ${}^\text{c}R_\text{i}$ and  ${}^\text{c}R_\text{o}$). Thus, a reduction of the outer sphere radius to 
\begin{equation*}
     {}^\text{s}R_{\text{o}} = {}^\text{s}R_{\text{i}}\left( \frac{^\text{c}R_{\text{o}}}{^\text{c}R_{\text{i}}} \right) ^{3/4}
\end{equation*}
makes both fields equal. If we now cut a long cylindrical shell  into two semi-infinite ones,  the field at the end is reduced by a factor 1/2. The same is true for a sphere cut into 2 hemispheres. Using a hemisphere as an end cap for the cylinder, the superposition of both fields  retains the original strength of the infinite cylinder here. This argument is strict only at the cutting plane of the cylinder, it does not apply for every distance from that plane.  Moreover,  it only represents an upper bound as a finite-length Halbach cylinder produces weaker fields. Depending on the length of the cylindrical section, the outer diameter of the hemisphere can be adapted to minimize the end effects. The corresponding correction factors are given in \cite{Soltner2023}, Appendix C. The design of discrete approximations of  a spherical geometry needs further modifications as discussed below. 

The opening angle for Halbach cylinders has been discussed in Ref.~\cite{CUFF2011}. Their approach is based on the fact that the total force is dominated by nearest-neighbor interactions in the limit of an infinitely thin cylinder. This allows the force-free opening angle to be determined in two limiting cases: the infinitely long and the very short cylinder. 

For infinitely long cylinders the force between two parallel line dipoles is proportional to $2\cos^2\beta -1$, where $\beta$ is the angle between their magnetic moments and their separation vector. The force vanishes at $\beta_0\!=\!45^\circ$, which is the analog of the magic angle for the line dipoles. 
Following the idea behind Eq.~(5) of Ref.\cite{CUFF2011}, one gets a force-free opening angle of $\theta_0\!=\!45^\circ$ in this case of an infinitely long cylinder.

\begin{figure*}[!htbp]
\centering
\includegraphics[width=\textwidth]{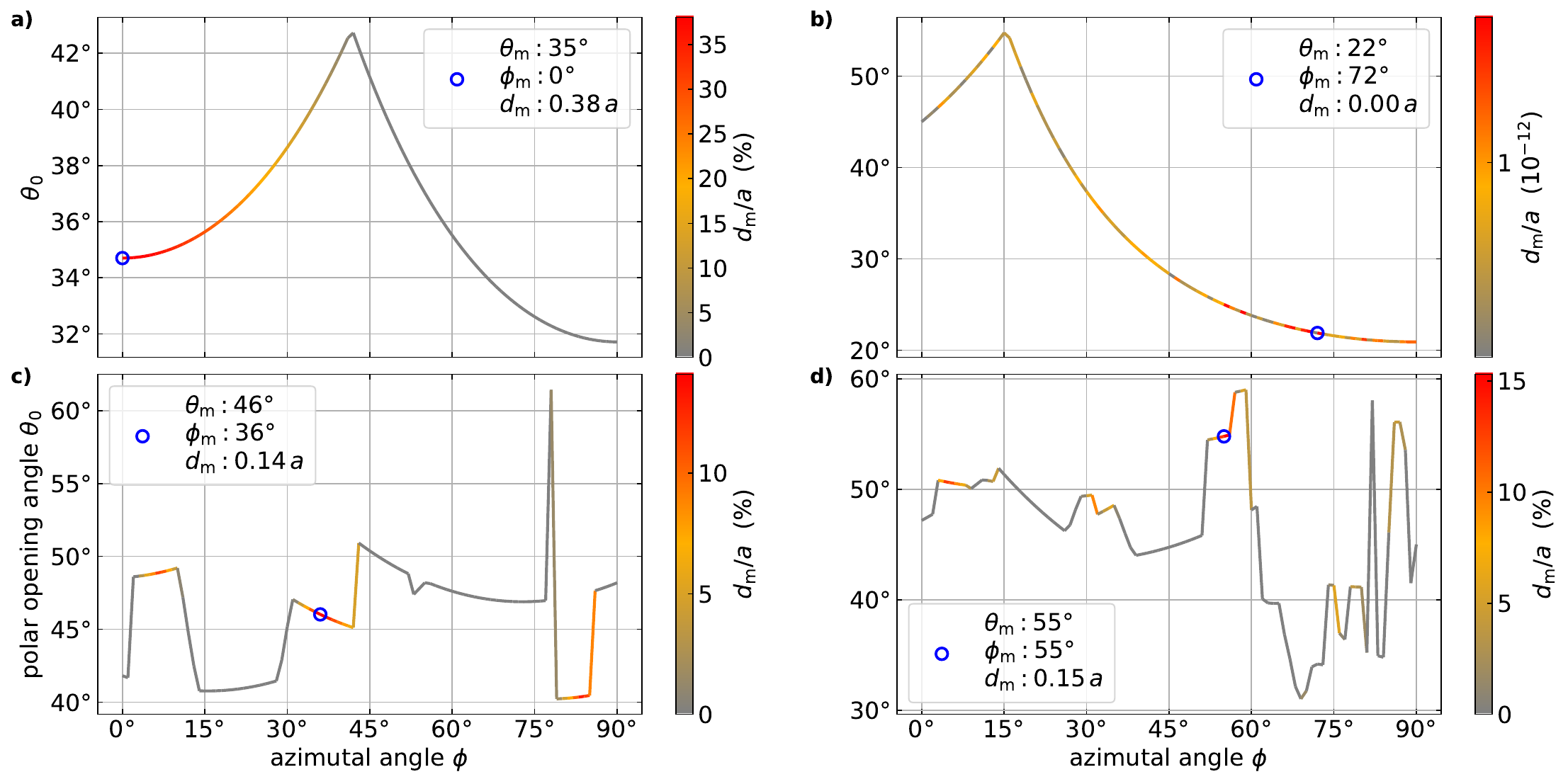}
\caption{The optimal opening angle for four magnet configurations: (a)~icosahedron, (b)~dodecahedron, (c) truncated icosahedron, (d)~truncated icosidodecahedron. The cutting angle $\theta_0(\phi)$ is shown as a colored line. The color provided in the color bar indicates the distance $d_\text{m}$ from the cutting plane to the nearest magnet. The blue circle indicates the optimal choice, with the corresponding numbers in the legend.}
\label{fig:5}
\end{figure*}

For an infinitely short cylinder (a ring), the force between parallel point dipoles is proportional to $3\cos^2\beta -1$. Following the arguments in \cite{CUFF2011}, the opening angle is $\theta_0\!=\!\theta_\text{m}\approx \!54.7^\circ$, which according to Fig.~\ref{fig:5} is the same as that obtained for homogeneous spherical shells.

Combining both arguments, one can conclude that the finite size cylinder will have a force-free opening angle in the range $45^\circ\!<\!\theta_0\! < \!54.7^\circ$. That range could be matched by a Fibonacci hemisphere, provided that more than $N\approx\left(127/9.7\right)^2/2\approx 150$ dipoles are used. The truncated icosahedron discussed in Fig.~\ref{fig:5} with its optimal opening angle of $46^\circ$ would also fit this range nicely with only 30 magnets.

This theoretical framework of continuously magnetized magnetic surfaces outlines a design space for permanent-magnet assemblies that balances mechanical accessibility with the magnetic field quality required for high-resolution magnetic resonance applications. We consider it as a useful guide for designs using discrete magnets. 

\begin{figure*}[!htbp]
\centering
\includegraphics[width=.95\textwidth]{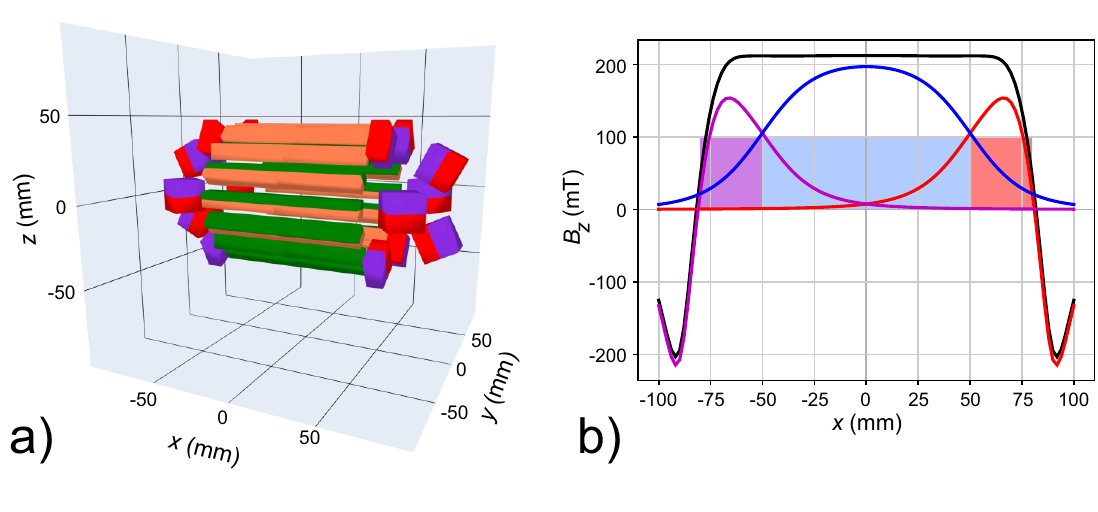}
\caption{Magpylib simulation of a spherocylindrical magnet assembled from cuboidal magnets: (a) The magnet consists of a central Halbach cylinder (magnets with north poles in orange and south poles in green), discretized into 16 cuboidal magnets (10 mm side length, 100 mm total length) arranged with their centers on a radius of 40 mm, and two icosahedral hemispherical end caps (magnets north in red, south in blue). Each end cap is constructed from four full cubic magnets (20 mm side length) and four half-cubes, with magnet centers located on hemispheres of radius 40 mm. The resulting magnetic field in the inside is oriented along the $z$-direction. (b) Axial magnetic field component $B_z$  evaluated along the central $x$-axis. Contributions from the icosahedral end caps (red/magenta), the Halbach cylinder (blue), and the combined structure (black) are shown. Colored background rectangles indicate the axial extent or open inner volume of each substructure, using the same color coding.}
\label{fig:6}
\end{figure*}

\subsection{Practical implications}
In principle, there is a wide range of possible cutting heights $h$ for a force-free opening. Choosing the largest radius of the opening disk by setting $h\!=\!0R$ seems to be the most practical, both in terms of interior accessibility and mechanical stability. It must be stressed that for the so called force-free opening angle $\theta_\text{0}$ there are shear forces, which must be accommodated by appropriately robust hinges.

Furthermore, and more importantly, as noted in the Introduction, magnetic materials with such geometries and continuously varying magnetization directions are not available. Therefore, cylindrical and spherical Halbach magnets must be approximated by using discrete segments with uniform magnetization, each positioned and oriented appropriately. In practice, condition $F\!=\!0$ will not be met exactly. However, even partial force compensation can reduce opening forces by one or two orders of magnitude, significantly enhancing mechanical safety and usability while preserving magnetic field strength and homogeneity.

When the opening plane overlaps with a magnet, one can make use of the possibility to turn the cutting plane around the $z$-direction by an azimuthal angle $\phi$, as discussed in Appendix G in \cite{Soltner2023} for rotational symmetric arrangements. However, there is no strict rotational symmetry for spherical arrangements of discrete magnets, so each configuration needs individual analysis. We present such a study for four different geometries in Fig.~\ref{fig:5}.

For this illustration, we have chosen the icosahedron formed by 12 magnets at its vertices, the dodecahedron (20 magnets), the truncated icosahedron (60 magnets) and the truncated icosidodecahedron (120 magnets). These configurations have icosahedral symmetry, and consequently their central magnetic field forms a saddle of fourth order \cite{RehbergBluemler2026}. To find the optimal opening angle, we restricted the investigation to $h\!=\!0R$, because this cut offers the largest opening aperture. For the normal $\hat{\mathbf{n}}$ of the cutting plan, we are now left with two parameters due to the lack of circular symmetry, the polar angle $\theta$ and the azimuthal angle $\phi$. In Fig.~\ref{fig:5} the force-free opening angle $\theta_0$ is determined as a function of $\phi$. This function $\theta_0(\phi)$ is plotted in different colors, where the color indicates the minimal distance from the cutting plane to the nearest magnet $d_\text{m}$. It is normalized  here by the common edge length $a$ of the Platonic or Archimedean solid, i.~e., the distance between neighboring magnets. To achieve maximum mechanical stability of the plastic frame while ensuring secure fixation of the magnets bonded to it,  $d_\text{m}$ should be maximized. That maximum is indicated by the red color and highlighted by a blue circle.

For the icosahedron characterized in Fig.~\ref{fig:5}a one obtains $\theta_0\!=\!34.7^\circ$ -- the value already introduced in the discussion of Fig.~\ref{fig:2} -- for an azimuthal angle of $\phi_\text{m}\!=\!0^\circ$. In that case, the magnet closest to the cutting plane has a distance of $d_\text{m}\approx 0.35 a$.

The dodecahedron elucidated in Fig.~\ref{fig:5}b brings a negative surprise in the pursuit to find an optimal cutting position: For any $\phi$, the cutting plane for a force-free opening touches a magnet, as indicated by the scale for $d_\text{m}$, which is of order $10^{-12}$ here. This small number is surely just a numerical artifact without any physical meaning. The fact that the cutting plane hits a magnet does not make it impossible to build such an apparatus, but it would certainly be a nuisance for the construction and stability of the magnet holding frame.

The geometry of the truncated icosahedron (the soccer ball shape), a magnet configuration introduced in \cite{RehbergBluemler2026} and elucidated  in Fig.~\ref{fig:5}c, seems to offer a realistic change to realize a force-free opening, with its maximal distance $d_\text{m}\!=\!0.14a$, which is obtained at an opening angle $\theta_0\!=\!46^\circ$ and the azimuthal angle $\phi\!=\!36^\circ$.

The same is true for the truncated icosidodecahedron with its 120 magnets, see Fig.~\ref{fig:5}d. For this configuration, it is noticeable that it contains large holes of decagon shape, which make it comparatively easy to reach the interior anyway. These large openings might reduce the need to open the whole construction when compared to the truncated icosahedron.

It should be noted that the tendency of the cutting angle to increase with the number of magnets, $N$, is also evident in the study shown in Fig.~\ref{fig:5}, consistent with the behavior observed for Fibonacci structures as $N$ increases. Furthermore, Fig.~\ref{fig:5} demonstrates that, due to the discretized nature of the magnetic structures, the resulting forces and feasible opening planes must be determined through numerical simulations of the magnet design.

The same applies to composite configurations. For example, the magnetic field inside a spherocylinder can be understood as the superposition of a central Halbach cylinder and two hemispherical end caps, as discussed in Fig.~\ref{fig:4}. However, the number, positions, and dimensions of the individual magnets must be optimized to balance field contributions and manufacturability using available magnet sizes (see Fig.~\ref{fig:6}). The optimization  is carried out here with the Magpylib simulation tool~\cite{Ortner2020}, and its result is illustrated in  Fig.~\ref{fig:6}b.

\section{Experimental Design and Results}
It should be noted that the two prototypes presented in the following were developed primarily to validate the concept of actuating magnetic opening and closing mechanisms with minimal force. To enable rapid fabrication, the systems were constructed using readily available, low-cost permanent magnets with short delivery times; however, these components exhibited limited reproducibility. In addition, the quantity procured was only sufficient for system assembly, precluding any selection or optimization from a larger pool of magnets. Consequently, the achievable magnetic field strength and homogeneity were not fully optimized.

\subsection{Icosahedron}
\begin{figure*}[!htbp]
\centering
\includegraphics[width=\textwidth]{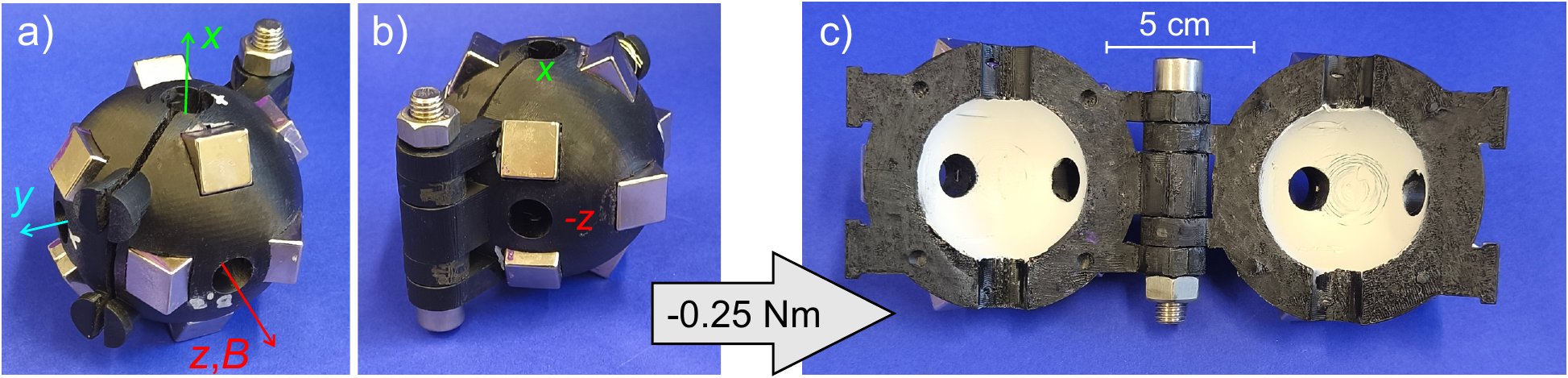}
\caption{Hinged icosahedral magnet composed of twelve NdFeB cubes (side length 20 mm with their centers at a radius of 40.5 mm), cutting plane defined by $\theta \!=\! 34.7^\circ$, $\phi\!=\!0^\circ$, and $h\!=\!0$.
(a) and (b)  Closed configuration shown from two different perspectives. The hinge is realized by an M8 screw from stainless steel. Two bollards allow the magnet to be secured. The measured opening torque is \qty{-0.25}{\newton\metre}. The coordinate system is overlayed (cf. Fig.~\ref{fig:2}).
(c) Open configuration. The surface of the inner (sample) volume ($R_{\text{i}} \!=\! 28$~mm) is colored white  for improved visibility. Two of the six openings ($\diameter$ \!=\!  15~mm)  are visible here.}
\label{fig:7}
\end{figure*}

The first prototype, shown in Fig.~\ref{fig:7}, is a Halbach sphere approximated by an icosahedral arrangement, a variation of the magnet assembly discussed  in detail in Ref.\cite{RehbergBluemler2026}. This assembly can be opened by means of the hinge located between the brass nuts. The hinge position was designed to open a cutting plane defined by  $\theta\!=\!\theta_{\text{m}}$, $\phi\!=\!90^\circ$,  and $h\!=\!0$. Its pivot point is located at a distance $r_\text{h}\!=\!59\text{ mm}$ from the center of the icosahedron.

\begin{figure*}[!hbtp]
\centering
\includegraphics[width=0.95\textwidth]{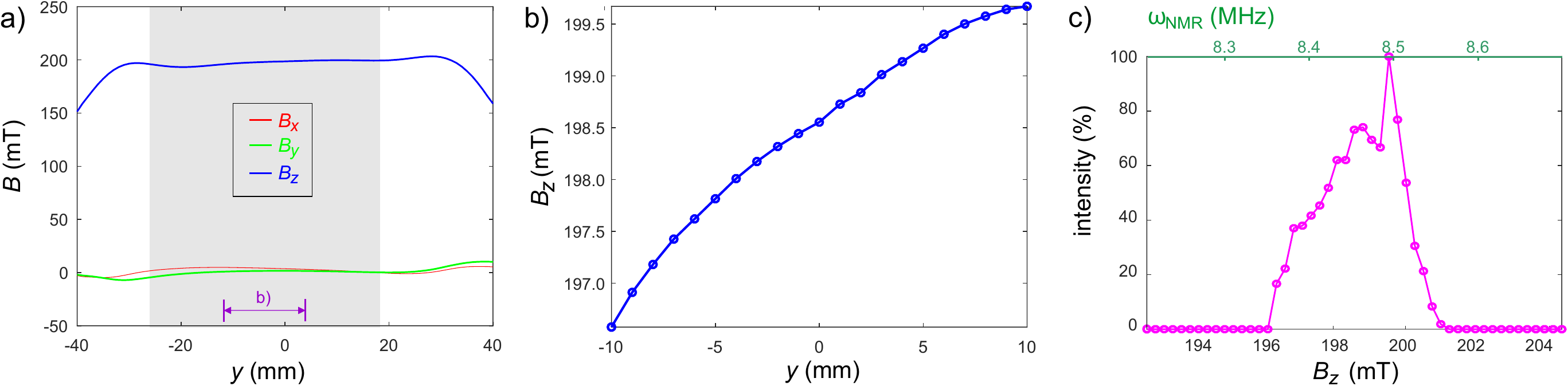}
\caption{Magnetic field of the icosahedron shown in Fig.~\ref{fig:7}, measured with a scanning Hall probe: (a) all three field components along the $y$-axis. Note that the inner opening of the shell has a diameter of only 56~mm as indicated by the gray background. (b) $B_z$ component in a zoomed region spanning 20~mm. (c) Field distribution for a central cylindrical volume (8~mm diameter, 20~mm length). The green scale on top refers to \textsuperscript{1}H-NMR frequencies.}
\label{fig:8}
\end{figure*}

The assembly consists of 12 permanent cubic magnets (purchased from Magnethandel item \#3982, Dortmund, Germany; dimensions $(20~\text{mm})^3$; Nd\textsubscript{2}Fe\textsubscript{14}B, grade N45, specified remanence $B_{\text{R}} \!=\! 1.33$–$1.36~\text{T}$). From the magnitude of the measured magnetic field $|B| \!=\! 196.5~\text{mT}$ (Fig.~\ref{fig:8}), an effective remanence of $B_{\text{R}} \!=\! 1.316~\text{T}$ was obtained by fitting.

The 12 magnets were bonded into two 3D-printed supports with epoxy glue such that their centers lay on a sphere of radius 40.5~mm, resulting in an inner free volume with radius $R_{\text{i}} \!=\! 28$~mm and a mass of 1040~g (see Fig.~\ref{fig:7}a-c). Additional access for field measurement in the closed state is provided by three 15 mm diameter through-holes, yielding six openings aligned along the Cartesian axes.

The torque $\tau$ needed to close the magnet was measured using a spring scale yielding $\tau\!=\! \qty{-0.25}{\newton\metre}$. The negative sign means that the force is attractive, i.e., the shell has a weak tendency to close by itself.

For a quantitative comparison, we need the force in the units given in Eq.~\ref{eq:3_force_unit}, which for the parameters $\theta\!=\!34.7^\circ$, $\phi\!=\!0^\circ$, and $h\!=\!0$ is calculated to be 
\begin{equation*}
F\approx-2.919\cdot10^{-5}\frac{\mu_0 m_\text{h}^2}{4\pi R^4}\ .
\end{equation*}
which means that we are more than four orders of magnitude below the maximum here, where according to Fig.~\ref{fig:2} the maximum is 0.6 in the same units.

To calculate the torque with respect to the hinge with its distance $r_\text{h}$ from the center of the icosahedron, we use the general formula 
\begin{equation}
\boldsymbol{\tau}_{O'}
=
\boldsymbol{\tau}_{O}
-
\left(\mathbf{r}_{O'}-\mathbf{r}_{O}\right)
\times
\sum_i \mathbf{F}_i\ ,
\end{equation}
where $\boldsymbol{\tau}_{O}$ is the torque with respect to the origin, here the center of the icosahedron. This torque $\boldsymbol{\tau}_{O}$ turns out to be zero for this configuration (and in general to all Halbach spheres with icosahedral symmetry). The closest point of the hinge pin, with its orientation along the direction $(-\sin\phi, \cos\phi,0)$, is located at the position
\begin{equation*}
\mathbf{r}_{O'}=r_\text{h}
\begin{pmatrix}
\cos\theta\,\cos\phi\\
\cos\theta\,\sin\phi\\
-\sin\theta
\end{pmatrix},
\end{equation*} 
a vector perpendicular to $\mathbf {\hat{n}}$.
The scalar opening force $F$ introduced above is the force component along  $\mathbf {\hat{n}}$ 
\begin{equation}
F=\mathbf {\hat{n}} \cdot \sum_i \mathbf{F}_i \ .
\end{equation}
Thus, it is the force component perpendicular to $\mathbf{r}_{O'}$, which simplifies the calculation of $\tau_\text{h}$, the torque component of the vector $\boldsymbol{\tau}$ along the direction of the hinge pin.  In SI units, according to the experimental parameters given above, this component then reads:
\begin{equation*}
\begin{split}
\tau_\text{h} &=F r_\text{h}  \\ &=2.919\cdot10^{-5} \frac{10^7}{(4\pi)^2}\frac{(6\cdot0.02^3 1.316)^2}{0.0405^4} \qty{0.059}{\newton\metre} \\& \approx \qty{-0.0027}{\newton\metre}
\end{split}
\end{equation*}

For the interpretation of this result, an error estimate is necessary. The angle can only be determined within a $\pm1^\circ$-range. This corresponds to a torque range of \qty{-0.049}{\newton\metre} to \qty{+0.049}{\newton\metre}.

The measured torque is found to be outside this range, which could be due to the fact that the calculation is based on point dipoles (or spheres), but not on cubes. Cubes with their tips bring parts of their volume closer to their neighbor, and due to the fact that forces scale with the inverse fourth power of the distance, these parts tend to raise the average force.

The value of the measured torque of \qty{-0.25}{\newton\metre} can be compared to the maximum torque expected for this arrangement, which is \qty{3.3}{\newton\metre}.
Thus, the reduction obtained here is more than one order of magnitude.

The magnetic field inside both Halbach magnets investigated here was measured using a three-axis Hall probe (MV2, Metrolab Technology SA, Plan-les-Ouates, Switzerland) mounted on a custom-built three-dimensional positioning stage. The stage is driven along the $x, y$ and $z$ axes by three stepper motors, which are controlled by an Arduino Mega R3 microcontroller using custom-developed software.

Figure~\ref{fig:8}a shows the three magnetic field components along the $z$-axis for the icosahedral assembly depicted in Fig.~\ref{fig:7} in its closed configuration. The field homogeneity is moderate, likely approaching the best achievable given the relatively low grade of the magnets and the limited precision of the 3D printing process. In particular, the acrylonitrile butadiene styrene (ABS) material used for printing exhibited slight deformation (approximately 1 mm) under magnetic forces in this configuration.

Despite this, the central region (Fig.~\ref{fig:8}b) displays reasonable homogeneity along the same axis. The homogeneity over a larger volume can be inferred from the magnetic field distribution (histogram) shown in Fig.~\ref{fig:8}c. This distribution is based on nearly 2000 Hall probe measurements acquired within a sampled central cylindrical volume (diameter 8 mm, length 20 mm). For convenience, the field values are additionally expressed as \textsuperscript{1}H NMR frequencies on a secondary (green) axis. The normalized peak exhibits a full width at half maximum (FWHM) of approximately 2.4 mT (102 kHz), corresponding to about 12000 ppm.

\subsection{Spherocylinder}
\begin{figure*}[!htbp]
\centering
\includegraphics[width=\textwidth]{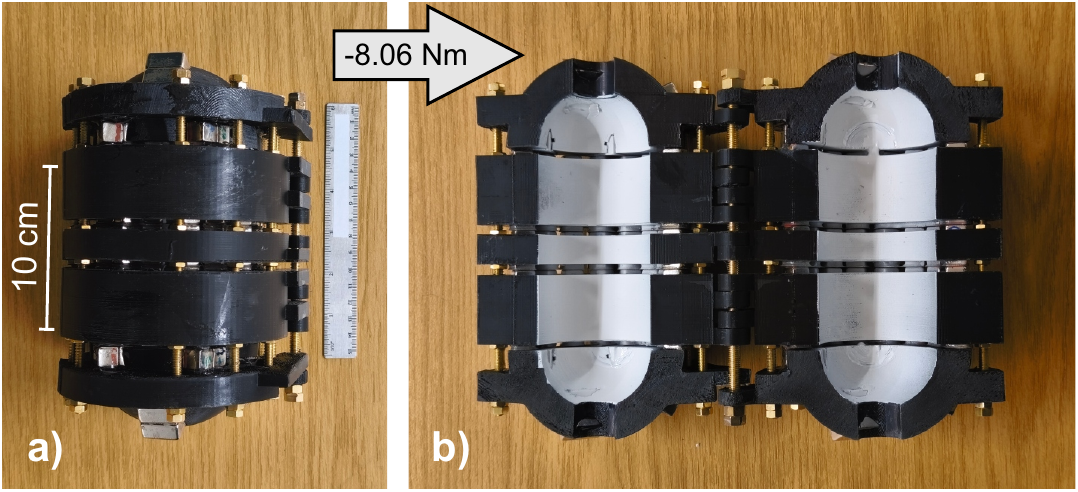}
\caption{Hinged spherocylinder composed of a 100~mm long Halbach cylinder and two hemispherical halves of an icosahedron, openable to an angle defined by $\theta = 33^{\circ}$ and $\phi = 90^\circ$. (a) Closed configuration, with the hinge located on the right (realized using an M6 threaded brass rod). The measured opening torque is \qty{-8.06}{\newton\metre}.
(b) Open configuration. The surface of the inner sample volume ($R_{\text{i}} = 28$~mm) is shown in white for improved visibility. The openings at the top and bottom are bisected by the opening plane.}
\label{fig:9}
\end{figure*}

In addition, a spherocylinder was constructed following the idea described in Fig.~\ref{fig:4}. The apparatus is shown in Fig.~\ref{fig:9}. The cylindrical section was designed to consist of 16 cuboid magnets of dimensions $100\times10\times10~\text{mm}^3$, magnetized parallel to one of the 10~mm sides. Since magnets of this size were not commercially available, each cuboid was assembled from two $40\times10\times10~\text{mm}^3$ magnets and one $20\times10\times10~\text{mm}^3$ magnet (Magnethandel items \#3860 and \#3977, Dortmund, Germany). The magnets were arranged with the shorter magnet positioned centrally and flanked on either side by the longer magnets.
The magnets were glued into 3D-printed supports so that the center of each cuboid lies on a circle of radius $R\! =\! 40.1$~mm, resulting in a cylindrical cavity with an inner radius of $R_\text{i}\! =\! 28$~mm. The outer diameter of the support cylinder is 120~mm.

The icosahedral hemispheres attached at each end of the cylinder contain two types of magnet: four $(20~\text{mm})^3$ cubes and four $20\times20\times10~\text{mm}^3$ cuboids. The cubic magnets were assembled from two $20\times20\times10~\text{mm}^3$ magnets (magnetized parallel to the 10~mm side; Magnethandel item \#3445). The $20\times20\times10~\text{mm}^3$ magnets, magnetized along the 20~mm side, were not available as stock items and were therefore constructed from four $(10~\text{mm})^3$ cubes each (obtained from Maqna, Germany). All magnets in the assembly had the same listed magnetic properties (Nd\textsubscript{2}Fe\textsubscript{14}B, grade N45, specified remanence: $B_{\text{R}} \!= \!1.33~\text{--}1.36~\text{T}$).

These magnets were glued into 3D-printed supports, again with their centers positioned on a circle of radius $R\! =\!  40.1$~mm, yielding an inner radius of $R_\text{i}\!  = \! 28$~mm. An opening with a diameter of 20~mm was left along the cylinder axis ($x$). The cylindrical section and the hemispheres were connected using 12 threaded brass rods of 5~mm diameter, together with a 6~mm rod serving as a hinge. The completed assembly has a total mass of 3.18~kg.

The opening angle was set at $\theta = 33^{\circ}$, placing it between the next pair of magnets in the cylinder.  The torque required to open the spherocylinder was measured accordingly to be \qty{-8.06}{\newton\metre}. The fact that the force is attractive is understandable, because the deliberately chosen angle $\theta\!  =\!  33^{\circ}$ is well below the range of force-free opening angles for cylinders ($45^\circ\text{--} 54.7^\circ$). The influence of the caps is expected to give a weak repelling force at this angle, which is clearly too small to compensate for the attractive force between the cylinder halves.

\begin{figure*}[!htbp]
\centering
\includegraphics[width=.95\textwidth]{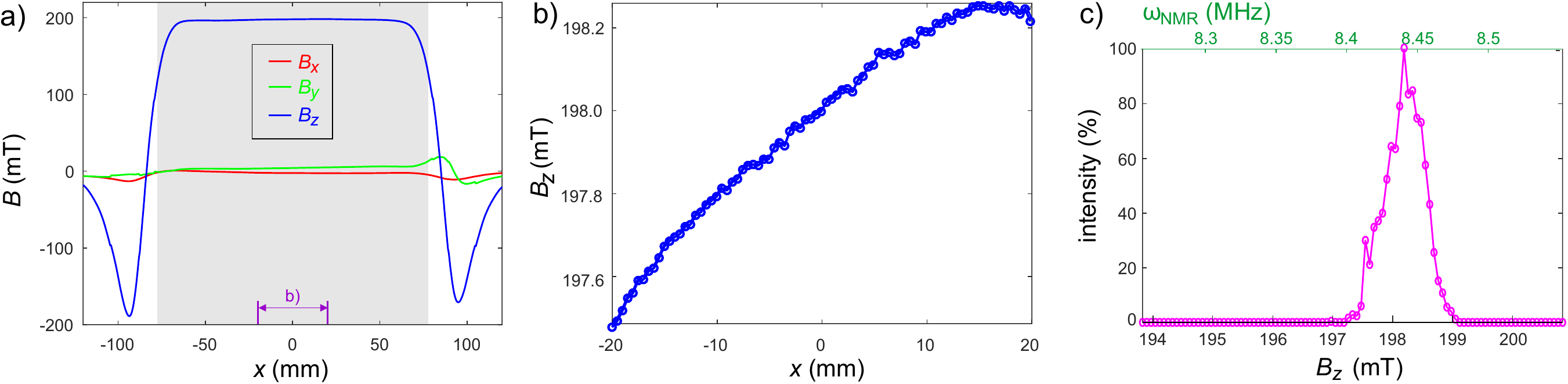}
\caption{ Magnetic field of the spherocylinder shown in Fig.~\ref{fig:9} (same representation as in Fig.~\ref{fig:8}): (a) all three field components along the $x$-axis. The inner accessible volume extends to 78~mm as indicated by the gray area. (b) $B_z$ component zoomed into the central region covering a distance 2 times larger than in Fig.~\ref{fig:8}b. (c) Field distribution calculated for a central cylindrical volume with a diameter of 10~mm and a length of 20~mm.}
\label{fig:10}
\end{figure*}

The same measurements and data processing as described for the icosahedron were applied to the magnetic field distribution within the closed spherocylinder shown in Fig.~\ref{fig:9}. The results are presented in Fig.~\ref{fig:10}. The field profile is considerably more symmetric than that of the icosahedral configuration shown in Fig.~\ref{fig:8}, the homogeneity inferred from the field distribution in Fig.~\ref{fig:10}c -- although evaluated over a slightly larger cylindrical volume (diameter 10~mm, length 20~mm) -- is approximately three times better (FWHM $\approx 0.8~\text{mT}$ or 35~kHz, corresponding to approximately 4100 ppm). A possible explanation for this is that there were more opportunities to choose magnetic blocks when optimizing the spherocylinder.

\section{Conclusion and Outlook}
In this work, we have developed and experimentally validated a general framework for the design of mechanically accessible spherical Halbach magnets that can be opened with minimal or vanishing force while preserving high magnetic field homogeneity. Starting from a dipolar Halbach sphere composed of discrete magnetic elements, the magic angle was identified as the asymptotic angle of the cutting plane. Selecting this plane with the help of a hinge eliminates the tensile forces across the interface, leaving only shear components that can be easily managed by the appropriate mechanical design.

The theoretical predictions were confirmed using prototype icosahedral assemblies, which demonstrated that the required opening forces can be reduced by orders of magnitude without significantly compromising the quality of the magnetic field. Residual asymmetries observed in the field measurements were attributed primarily to practical limitations such as minor geometric misalignments and fabrication tolerances, rather than fundamental constraints of the concept, hence more carefully built systems should also result in better homogeneities.

The extension of this approach to spherocylindrical geometries further demonstrates its versatility while enhancing mechanical accessibility and scalability to larger volumes. In particular, combining Halbach cylinders with hemispherical end caps enables the generation of highly homogeneous magnetic fields over expandable volumes with exceptionally efficient use of magnetic material in terms of both mass and volume. At the same time, it preserves the key advantage of allowing controlled opening with manageable forces.

Overall, the presented concept addresses a long-standing challenge in permanent magnet design: combining excellent field homogeneity with full mechanical accessibility. The results open new possibilities for compact, energy-efficient magnetic resonance systems and other applications requiring well-defined magnetic fields in enclosed yet accessible volumes.

Future work may focus on optimized magnet selection, improved assembly precision, and extension to higher-order multipole configurations to further enhance performance.

\bibliography{references.bib}

@article{Bauer2009,
Author = {Bauer, C and Raich, H and Jeschke, G and Blümler, P},
Title = {Design of a permanent magnet with a mechanical sweep suitable for
   variable-temperature continuous-wave and pulsed EPR spectroscopy},
Journal = {Journal of Magnetic Resonance},
Year = {2009},
Volume = {198},
Number = {2},
Pages = {222-227},
Month = {JUN},
DOI = {10.1016/j.jmr.2009.02.010},
}

@incollection{BluCasa2015,
   author = {Blümler, P. and Casanova, F.},
   title = {Hardware Developments: Halbach Magnet Arrays},
   booktitle = {Mobile NMR and MRI: Developments and Applications},
   editor = {Johns, M. and Fridjonsson, E. O. and Vogt, S. and Haber, A.},
   publisher = {Royal Chemical Society},
   address = {Cambridge},
   chapter = {4},
   pages = {133-157},
   DOI = {10.1039/9781782628095-00133},
   year = {2015},
   type = {Book Section}
}

@article{ Chen2007,
Author = {Chen, Jizhong and Xu, Chunyan},
Title = {Design and analysis of the novel test tube magnet as a device for
   portable nuclear magnetic resonance},
Journal = {IEEE Transactions on Magnetics},
Year = {2007},
Volume = {43},
Number = {9, 1},
Pages = {3555-3557},
Month = {SEP},
DOI = {10.1109/TMAG.2007.901888},
ISSN = {0018-9464},
}

@article{CUFF2011,
   author = {Windt, Carel W. and Soltner, Helmut and van Dusschoten, Dagmar and Blümler, Peter},
   title = {{A portable Halbach magnet that can be opened and closed without force: The NMR-CUFF}},
   journal = {Journal of Magnetic Resonance},
   volume = {208},
   number = {1},
   pages = {27-33},
   ISSN = {1090-7807},
   DOI = {10.1016/j.jmr.2010.09.020},
   year = {2011},
   type = {Journal Article}
}

@book{Furlani2001,
  address = {San Diego},
  author = {Furlani, E. P. },
  isbn = {0-12-269951-3.},
  publisher = {Academic Press},
  title = { Permanent Magnet and Electromechanical Devices: Materials, Analysis, and Applications},
  DOI = {10.1016/B978-012269951-1/50004-8},
  year = {2001}
}

@article{Halbach1980,
   author = {Halbach, K.},
   title = {{Design of permanent multipole magnets with oriented rare earth cobalt material}},
   journal = {Nuclear Instruments and Methods},
   volume = {169},
   number = {1},
   pages = {1-10},
   DOI = {10.1016/0029-554X(80)90094-4},
   year = {1980},
   type = {Journal Article}
}

@book{Johns2015,
    author = {},
    editor = {Johns, Michael L and Fridjonsson, Einar O and Vogt, Sarah J and Haber, Agnes},
    title = {Mobile NMR and MRI: Developments and Applications},
    publisher = {The Royal Society of Chemistry},
    year = {2015},
    month = {10},
    isbn = {978-1-84973-915-3},
    doi = {10.1039/9781782628095},
    url = {https://doi.org/10.1039/9781782628095},
}

@article{Leupold1993,
Author = {Leupold, HA and Tilak, AS and Potenziani, E},
Title = {Adjustable Mutli-Tesla Permanent Magnet Field Sources},
Journal = {IEEE Transactions on Magnetics},
Year = {1993},
Volume = {29},
Number = {6, 1},
Pages = {2902-2904},
Month = {NOV},
Note = {1993 IEEE International Magnetics Conference (INTERMAG 93), Stockholm,
   Sweden, Apr 13-16, 1993},
Organization = {IEEE, MAGNET SOC},
DOI = {10.1109/20.281092},
ISSN = {0018-9464},
}

@article{Leupold1994,
Author = {Leupold, HA and Potenziani, E and Tilak, AS},
Title = {LIGHTWEIGHT, DISTORTION-FREE ACCESS TO INTERIORS OF STRONG
   MAGNETIC-FIELD SOURCES},
Journal = {Journal of Applied Physics},
Year = {1994},
Volume = {76},
Number = {10, 2},
Pages = {6856-6858},
Month = {NOV 15},
Note = {6th Joint Magnetism and Magnetic Materials-Intermag Conference,
   Albuquerque, NM, Jun. 20-23, 1994},
Organization = {AMER INST PHYS; IEEE, MAGNET SOC; MINERALS MET \& MAT SOC; AMER SOC
   TESTING \& MAT; USN, OFF NAVAL RES; AMER CERAM SOC},
DOI = {10.1063/1.358530},
ISSN = {0021-8979},
Unique-ID = {WOS:A1994PT84800274},
}

@article{Leupold2000,
Author = {Leupold, HA and Tilak, A and Potenziani, E},
Title = {Permanent magnet spheres: Design, construction, and application
   (invited)},
Journal = {Journal of Applied Physics},
Year = {2000},
Volume = {87},
Number = {9, 2},
Pages = {4730-4734},
Month = {MAY 1},
Note = {44th Annual Conference on Magnetism and Magnetic Materials, San Jose,
   CA, Nov. 15-18, 1999},
Organization = {Amer Inst Phys; Magnet Soc Inst Elect \& Electr Engineers; Minerals Met
   \& Mat Soc; Amer Soc Testing \& Mat; USN Off Res; Amer Ceram Soc; Amer
   Phys Soc},
DOI = {10.1063/1.373141},
ISSN = {0021-8979},
Unique-ID = {WOS:000086727200021},
}

@article{Ortner2020,
   author = {Ortner, M. and Bandeira, L.G.C. },
   title = { Magpylib: a free Python package for magnetic field computation},
   journal = {SoftwareX},
   volume = {11},
   pages = {100466},
   DOI = {10.1016/j.softx.2020.100466},
   year = {2020},
   type = {Journal Article}
}

@misc{patent_spheres,
  author       = {Bl{\"u}mler, P. and Soltner, H.},
  title        = {Halbach magnet for magnetic resonance that can be opened and closed without effort},
  howpublished = {European Patent EP2365353},
  year         = {2011},
}

@misc{Rehberg2026b,
  author       = {Rehberg, Ingo},
  title        = {{Dipole Cluster Inspector - A Python GUI for
                   Exploring Thousands of Magnetic Configurations
                  }},
  month        = jul,
  year         = 2026,
  publisher    = {Zenodo},
  version      = {2.2.0},
  doi          = {10.5281/zenodo.10084573},
  url          = {https://doi.org/10.5281/zenodo.10084573},
}

@article{RehbergBluemler2026,
  title = {{Discretized Halbach spheres: Icosahedral symmetry for optimal field homogeneity}},
  author = {Rehberg, Ingo and Bl\"umler, Peter},
  journal = {Phys. Rev. Appl.},
  volume = {25},
  issue = {5},
  pages = {054009},
  numpages = {20},
  year = {2026},
  month = {May},
  publisher = {American Physical Society},
  doi = {10.1103/hyv2-s2tf},
  url = {https://link.aps.org/doi/10.1103/hyv2-s2tf}
}

@misc{RehbergBluemler2026a,
  author       = {Rehberg, Ingo and Blümler, Peter},
  title        = {{Halbach\_two\_point\_oh: Optimize Uniform Fields with
                   Permanent Magnets}},
  month        = jun,
  year         = 2026,
  publisher    = {Zenodo},
  version      = {2.1.1},
  doi          = {10.5281/zenodo.20669385},
  url          = {https://doi.org/10.5281/zenodo.20669385},
}

@article{RehbergBluemler2025,
  title =
  {Analytic approach to creating homogeneous fields with finite-size magnets},
  author = {Rehberg, Ingo and Bl\"umler, Peter},
  journal = {Phys. Rev. Appl.},
  volume = {23},
  issue = {6},
  pages = {064029},
  numpages = {19},
  year = {2025},
  month = {Jun},
  publisher = {American Physical Society},
  doi = {10.1103/9nnk-jytn},
 }

@article{Soltner2023,
   author = {Soltner, H. and Blümler, P.},
   title = {{Practical Concepts for Design, Construction and Application of Halbach Magnets in Magnetic Resonance}},
   journal = {Applied Magnetic Resonance},
   volume = {54},
   pages = {1701-1739},
   DOI = {10.1007/s00723-023-01602-2},
   year = {2023},
   type = {Journal Article}
}

@article{Swinbank2006,
author = {Swinbank, Richard and James Purser, R.},
title = {Fibonacci grids: A novel approach to global modelling},
journal = {Quarterly Journal of the Royal Meteorological Society},
volume = {132},
number = {619},
pages = {1769-1793},
doi = {https://doi.org/10.1256/qj.05.227},
url = {https://rmets.onlinelibrary.wiley.com/doi/abs/10.1256/qj.05.227},
eprint = {https://rmets.onlinelibrary.wiley.com/doi/pdf/10.1256/qj.05.227},
year = {2006}
}

@article{Tretiak2019,
Author = {Tretiak, Oleg and Blümler, Peter and Bougas, Lykourgos},
Title = {Variable single-axis magnetic-field generator using permanent magnets},
Journal = {AIP Advances},
Year = {2019},
Volume = {9},
Number = {11},
DOI = {10.1063/1.5130896},
}

@article{Yung1998,
  author        = {Yung, K. W. and Landecker, P. B. and Villani, D. D.},
  title         = {An Analytic Solution for the Force Between Two Magnetic Dipoles},
  journal       = {Magnetic and Electrical Separation},
  year          = {1998},
  volume       = {9},
  number       = {1},
  pages        = {39--52},
  doi          = {10.1155/1998/79537},
}
\end{document}